\documentclass[aps, prd, showpacs, superscriptaddress, nofootinbib, twocolumn]{revtex4}
\usepackage{amssymb,amsmath,microtype}
\usepackage{graphicx}
\usepackage[normalem]{ulem}
\usepackage{color}

\begin{document}

\title{Searching for features of a string-inspired inflationary model with cosmological observations}

\author{Yi-Fu Cai}
\email{yifucai@ustc.edu.cn}
\affiliation{Key Laboratory for Research in Galaxies and Cosmology, Department of Astronomy,
University of Science and Technology of China, Chinese Academy of Science, Hefei, Anhui 230026, China}
\affiliation{Department of Physics, McGill University, Montr\'eal, Quebec H3A 2T8, Canada}

\author{Elisa G. M. Ferreira}
\email{elisa.ferreira@mail.mcgill.ca}
\affiliation{Department of Physics, McGill University, Montr\'eal, Quebec H3A 2T8, Canada}

\author{Bin Hu}
\email{hu@lorentz.leidenuniv.nl}
\affiliation{Institute Lorentz, Leiden University, PO Box 9506, Leiden 2300 RA, The Netherlands}

\author{Jerome Quintin}
\email{jquintin@physics.mcgill.ca}
\affiliation{Department of Physics, McGill University, Montr\'eal, Quebec H3A 2T8, Canada}

\begin{abstract}
The latest {\it Planck} results show a power deficit in the temperature
anisotropies near $\ell \approx 20$ in the cosmic microwave background (CMB).
This observation can hardly be explained within the standard inflationary
$\Lambda$-cold-dark-matter ($\Lambda$CDM) scenario. In this paper we consider a string theory
inspired inflationary model (axion monodromy inflation) with a step-like modulation in
the potential which gives rise to observable signatures in the primordial perturbations.
One interesting phenomenon is that the primordial scalar modes experience a sudden
suppression at a critical scale when the modulation occurs. By fitting to the CMB data,
we find that the model can nicely explain the $\ell \approx 20$ power deficit anomaly
as well as predict specific patterns in the temperature-polarization correlation and
polarization autocorrelation spectra. Though the significance of the result is not sufficient
to claim a detection, our analysis reveals that fundamental
physics at extremely high energy scales, namely, some effects inspired by string theory,
may be observationally testable in forthcoming cosmological experiments.
\end{abstract}


\maketitle

\section{Introduction}

With accumulated high-precision measurements of the cosmic microwave background (CMB)
radiation~\cite{Hinshaw:2012aka, Ade:2013zuv, Ade:2015xua}, it is believed that our
universe has experienced a dramatic phase of expansion at very early times as described by the
inflationary $\Lambda$CDM cosmology from which a power law spectrum of primordial
perturbations is obtained~\cite{Mukhanov:1990me}. The recently released
{\it Planck} data, however, reported deviations from a power law connected to a deficit
at multipoles $\ell \approx 20 - 40$ in the temperature power spectrum~\cite{Ade:2015lrj}.
Although this anomaly does not have sufficient statistical significance due to the cosmic
variance at these multipoles, it is of enough interest to ask whether such an experimental
signature provides a hint to new physics beyond $\Lambda$CDM.

In order to understand cosmological data from fundamental physics, we consider an important class of large field inflationary
cosmology realized by string theory, which is dubbed axion monodromy inflation (AMI)~\cite{Silverstein:2008sg, McAllister:2008hb}.
In the corresponding stringy setup, a number of axion fields coupled to fluxes can realize a super-Planckian field variation
with a soft shift symmetry breaking along the axion potential due to couplings or D-branes. This is how a class of monomial
potentials has been achieved~\cite{McAllister:2014mpa}. This model and its extensions have drawn a lot of
attention in the literature (e.g.,\ see~\cite{Flaugeretal, Cai:2014vua} and references therein).
In particular, it was observed in~\cite{Cai:2014vua} that axion monodromies can be obtained in terms of D3- and D5-branes
with torsional cycles in which there exists at least one scalar mode that is free from the dangerous moduli stabilization
effects from the K\"{a}hler potential. This scalar field, when applied to the early universe, can drive a sufficiently
long phase of inflation where the potential parameters are sensitive to flux couplings, namely, the potential experiences
a modulation at a critical value of the inflaton field. As a result, it provides an interesting implementation of
step inflation~\cite{Starobinsky:1992ts, Adams:1997de} from the perspective of string theory.

In this paper we aim at examining observational signatures of this type of inflationary model in cosmological surveys.
Specifically, we present an estimate of the power spectrum of primordial curvature perturbations and analytically find that
it possesses a suppression feature at a critical length scale. Using the {\it Planck} 2013 data, we perform a
numerical analysis and find that this feature can nicely interpret the anomaly at low multipoles as observed in the recent
CMB observations which cannot be explained by the standard model, i.e.\ $\Lambda$CDM.
Afterwards, we compute the E-mode polarization spectra, and interestingly, we find that our model also predicts
nontrivial patterns on these spectra. We take the convention of the {\it Planck} team throughout the paper.

\section{Model}
We work with the string-inspired inflationary model of axion monodromy in
the presence of torsional cycles and flux couplings~\cite{Cai:2014vua}.
In this framework, the 4-dimensional effective action is described by a
canonical scalar field with a polynomial potential coming from either the
Chern-Simons term or flux couplings. Due to a variation of the flux coupling,
the inflaton's potential can be expressed as
\begin{equation}\label{V_para}
 V(\varphi) = \lambda M_{\mathrm{pl}}^4 (\varphi/M_{\mathrm{pl}})^n {\cal F}_V(\varphi) \,,
\end{equation}
with
\begin{equation}\label{FV}
{\cal F}_V(\varphi) = 1+ {\xi^2}/[ 1 +e^{-2 c_H(\varphi^2-\varphi_c^2)/M_{\mathrm{pl}}^2}] \,,
\end{equation}
where $M_{\mathrm{pl}}$ is the reduced Planck mass (see~\cite{Cai:2014vua} for the string theory implementation with $n=2$).

The coefficient of the potential takes the form $(1+\xi^2) \lambda$ in the ultraviolet (UV) regime ($|\varphi| > \varphi_c$)
but becomes $\lambda$ in the infrared (IR) regime ($|\varphi| \leq \varphi_c$), where $\xi$ is a dimensionless parameter.
The other dimensionless parameter $c_H$ is associated with the smooth transition between the UV and IR regimes of the
potential near the critical scale characterized by $\varphi_c$. As was observed in~\cite{Cai:2014vua}, the inflaton field
undergoes standard slow roll dynamics in both the UV and IR regimes, but near the potential modulation,
the slow roll condition is briefly broken.

In this case it is convenient to introduce the generalized slow roll (GSR) parameters~\cite{Stewartetal},
which up to $i$-th order are given by $\epsilon \equiv -\frac{\dot{H}}{H^2}$, $\eta_i \equiv \frac{\varphi^{(i+1)}}{H^i\dot\varphi}$,
where a dot denotes a cosmic time derivative. Then one can follow the formalism of the GSR approximation
to derive an analytic expression of the power spectrum. Here we skip the detailed procedure (which will
be presented in a follow-up paper) but directly write down the power spectrum of primordial curvature perturbation as
\begin{eqnarray}\label{Pk_axion}
 \mathcal{P}_{\mathcal{R}}(k) = \bar{\mathcal{P}}_{\mathcal{R}}(k) {\cal F}_V(\varphi(k)) {\cal F}_M(k)~,
\end{eqnarray}
where $\bar{\mathcal{P}}_{\mathcal{R}}$ is the featureless power spectrum from the standard model and ${\cal F}_M$ represents
the modifications introduced by the GSR approximation. Similar to the standard case, the power spectrum can
be parametrized by $\bar{\mathcal{P}}_{\mathcal{R}} \equiv A_s ({k}/{k_{\ast}})^{n_s-1}$, where $A_s$ and $n_s$ are the
amplitude and the spectral index, respectively, and $k_\ast$ is the pivot scale.
For a specific power law potential, we have
\begin{align}
 \varphi(k) &\approx \varphi_\ast - n\frac{M_{\mathrm{pl}}^2}{\varphi_\ast} \ln\left(\frac{k}{k_\ast}\right)\,, \nonumber \\
 \varphi_\ast^2 &\approx \frac{n}{2}(4N_\ast+n) M_{\mathrm{pl}}^2\,,
\end{align}
and the power spectrum quantities $A_s$ and $n_s$ can be approximately written
as~\cite{Planck:2013jfk, Ade:2015lrj}
\begin{equation}\label{As_ns}
  A_s \approx \frac{\lambda}{12\pi^2 n^2} \left|\frac{\varphi_\ast}{M_{\mathrm{pl}}}\right|^{n+2} \,, \qquad n_s-1 \approx \frac{-2(n+2)}{4N_\ast+n}\,.
\end{equation}

To apply these relations it is convenient to introduce $k_c$ at $\varphi(k_c)=\varphi_c$ where the feature is located.
The factor ${\cal F}_M$ grasps the signatures brought by the potential modulation, and therefore,
its value equals unity away from $k_c$ but exhibits the GSR feature near $k_c$. For $\xi>1$,
these features have an oscillatory behavior near $k_c$, analogous to~\cite{Dvorkin:2009ne}
where a quadratic model with a different modulation form was considered.

In this paper we focus our interest in the case of $\xi < 1$. Accordingly, the GSR feature is very smooth and is only located
in the vicinity of the potential modulation and mainly depends on the evolution of $\epsilon$. To leading order one has
${\cal F}_M \simeq 1 +(4f_\alpha-2)\epsilon +2f_\alpha \eta_1 +(3f_\alpha^2+\frac{5\pi^2}{12}-4)\eta_1^2 +(\frac{\pi^2}{12}-f_\alpha^2)\eta_2$,
where $f_\alpha \equiv 2 - \ln 2 - \gamma$ with $\gamma$ being the Euler-Masheroni constant.
This factor can be further parametrized as follows,
\begin{equation}
 {\cal F}_M(k) \simeq 1 + \frac{ 9 c_H^2 \xi^2 \varphi^2 [e^{ \frac{2 c_H}{M_{\mathrm{pl}}^2} (\varphi^2-\varphi_c^2) } -1] }{M_{\mathrm{pl}}^2 \cosh[ \frac{2 c_H}{M_{\mathrm{pl}}^2} (\varphi^2-\varphi_c^2) ] [ e^{ \frac{2 c_H}{M_{\mathrm{pl}}^2} (\varphi^2-\varphi_c^2)} +1]^3 }~.
\label{FM}
\end{equation}

\begin{figure}
 \begin{center}
  \includegraphics[scale=0.81]{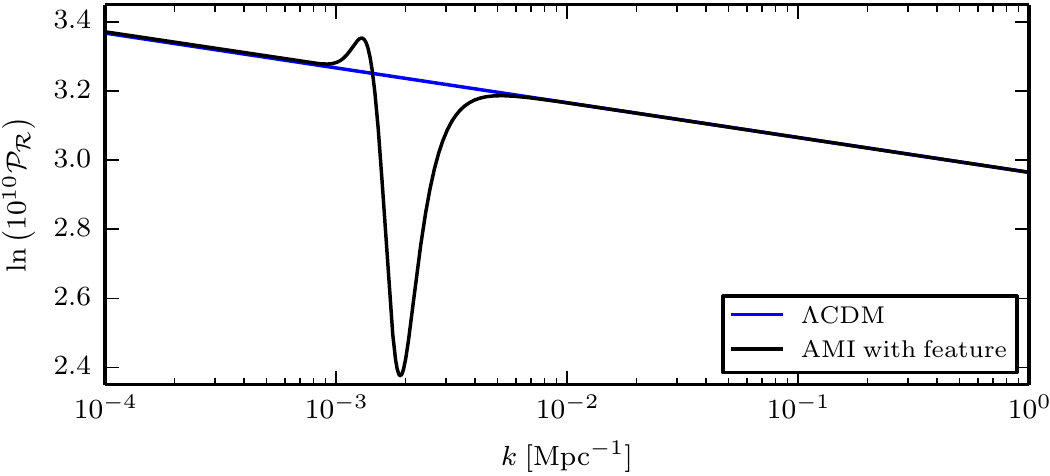}
  \caption{Power spectra of primordial curvature perturbations $\mathcal{P}_{\mathcal{R}}$ as a function of the comoving wave number $k$.
  The blue curve shows the standard $\Lambda$CDM model and the black curve presents an example of the
  AMI with feature given by~\eqref{Pk_axion}.}
  \label{fig:power_spectrum}
 \end{center}
\end{figure}

The parametrized form~\eqref{Pk_axion} together with Eqs.~\eqref{FV} and~\eqref{FM}
gives the primordial power spectrum of the AMI with feature, which is described by three
extra parameters on top of $A_s$ and $n_s$: $\xi$, $c_H$, and $k_c$ (or $\varphi_c$ equivalently),
as demonstrated in Fig.\ \ref{fig:power_spectrum}.
However, one extra parameter remains undetermined, i.e.\ the {\it e}-folding number ($N_\ast$)
at the pivot scale, since it involves detailed knowledge of
the reheating theory following it~\cite{Liddle:2003as, Martin:2010kz}.
A different value of $N_\ast$ leads to a different $\varphi_c$, altering the position of the feature.
This shows that $N_\ast$ is highly degenerate with, at least, $k_c$. In the literature,
$N_\ast$ is sometimes fixed for simplicity~\cite{Ade:2015lrj, Benetti:2012wu}, but this leads to preferred
values of $n$, as seen from Eq.~\eqref{As_ns}. For the sake of generality, we investigate both possibilities:
having $N_\ast$ fixed and having it as a free parameter.
Additionally, the parameter space can be further reduced by imposing theoretical constraints, namely,
the power spectrum must always be positive, i.e.\ ${\cal F}_M > 0$. We impose this
constraint in all numerical calculations below.

\section{Methodology}

We wish to compare the theoretical predictions of our model and to test the validity of our
parametrization of the power spectrum with the current data. We perform a global
fitting by running the CosmoMC package~\cite{Lewis:2002ah}, a Markov Chain Monte Carlo (MCMC) parameter sampler.
We modify the publicly available CAMB~\cite{Lewis:1999bs} Boltzmann code with
our parametrization~\eqref{Pk_axion}, which gives the power spectrum with a feature described
by $c_H$, $\xi$, and $k_c$. Those are combined with the ``vanilla'' parameters of {\it Planck}'s
one-parameter extension of the baseline $\Lambda$CDM model~\cite{Ade:2015lrj, Planck:2013jfk} consisting of 7 parameters
($\Omega_bh^2$, $\Omega_ch^2$, $\tau$, $\Theta_s$, $n_s$, $A_s$, $r$) where $\Omega_bh^2$ and $\Omega_ch^2$
are the baryon and cold dark matter densities, $\tau$ is the optical depth to reionization,
$\Theta_s$ is the ratio of the sound horizon at decoupling to the angular diameter distance
to the last scattering surface (multiplied by 100), $n_s$ is the spectral index, $A_s$
is the primordial amplitude, and $r$ is the tensor-to-scalar ratio.
As usual, we assume adiabatic initial conditions and a flat background universe.

Due to the limited knowledge of reheating, there remains uncertainty in
determining the exact $e$-folding number $N_{\ast}$ when the pivot
mode ($k_{\ast}$) crosses the horizon. In order to take this into account, we first take the {\it Planck}
convention~\cite{Planck:2013jfk} by setting $50<N_{\ast}<60$ and vary
$N_{\ast}$ when we perform the global fitting.
Then, in order to have a closer look at
the effect of fixing $N_{\ast}$ on the parameter estimation, we perform three extra runs.
Specifically, we fix $N_{\ast}$ at $50$, $55$, and $60$, which correspond to
$k_{\ast}\approx 1.0,~0.05$, and $0.002$ ${\rm Mpc}^{-1}$ according to the modeling of
entropy generation at the end of inflation given by Eq.~(24) in~\cite{Planck:2013jfk}.
In this case, when the details of the entropy generation process are fixed, the relation between $N_\ast$ and $k_\ast$ is known.
Thus, fixing $k_\ast$ fixes $N_\ast$ and vice-versa.
redHere we emphasize again, in the case of fixing $N_{\ast}$, we only investigate a specific formula for
reheating, the one that is adopted by the Planck collaboration~\cite{Planck:2013jfk}. As will be demonstrated below,
the sensitivity of the feature in different multipole ranges is expected to depend on the choice of $k_\ast$, so it is crucial to set
the relationship between $N_{\ast}$ and $k_{\ast}$ consistently. This is important in its own right, but the exploration of
different reheating processes is beyond the scope of this paper.

When considering $N_{\ast}$ as a free parameter,
we do not assume any specific reheating scenario. This means that the relation between $N_\ast$ and $k_\ast$ is not known.
In this case, the pivot scale $k_\ast$ is fixed at the value
$k_\ast=0.05~\mathrm{Mpc}^{-1}$, which is equivalent to ``marginalizing'' over the uncertainty of the reheating mechanism.
However, it is important to know $N_\ast$ since it changes the position of the possible feature,
and this is determined by the MCMC analysis.

We use the {\it Planck} 2013 {\it TT} power spectrum, both the low-$\ell$ ($2\le\ell<50$)
and high-$\ell$ ($50\le\ell\le2500$) multipole data. In order to break the degeneracy between
reionization optical depth and the primordial amplitude, we include the WMAP~\cite{Bennett:2012zja}
low-$\ell$ polarization power spectrum ($2\le\ell\le32$). Besides that, we adopt the baryon
acoustic oscillation data from BOSS ``LOWZ'' and CMASS-DR11 surveys~\cite{Anderson:2013zyy}
in order to obtain a better estimation on $\Omega_\mathrm{m}$ and $H_0$, {\it etc}. We name the above data
compilation {\it Planck}13 in the following analysis. Moreover, we implement a
logarithmic prior for $\xi$ and $k_c$, namely
$\log_{10}\xi\in[-8,0]$ and $\log_{10} (k_c/{\rm Mpc}^{-1}) \in[-4,0]$, to be able to
span a large parameter space. The theoretical constraint of $\mathcal{F}_M>0$ restricts
the value of $c_H$ to be close to unity. Given this observation, we choose a flat prior for $c_H$ such that $c_H\in[0,1]$.

\section{Results}

\begin{figure}
\begin{center}
\includegraphics[scale=0.265]{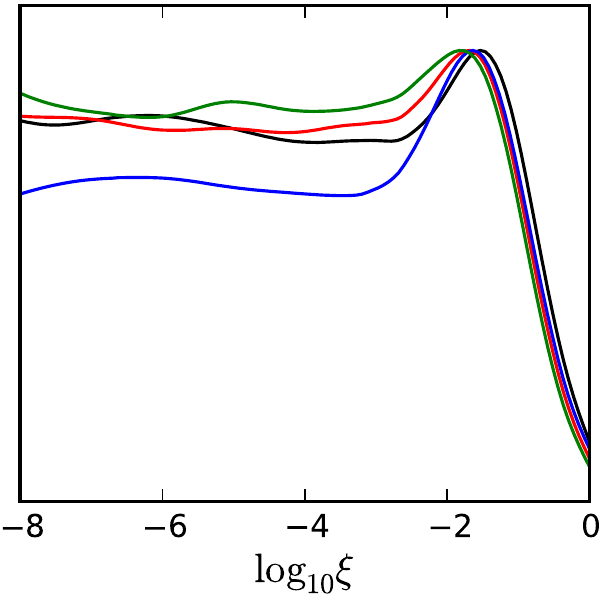}
\includegraphics[scale=0.265]{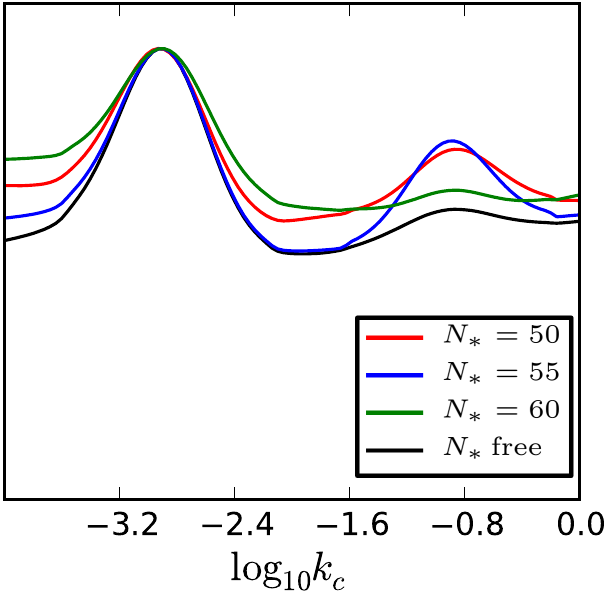}
\caption{One-dimensional posteriors of the feature parameters $\xi$ and $k_c$
for different {\it e}-folding numbers $N_{\ast}$ as a result of the fitting to
the {\it Planck}13 data.}
\label{figure:1D_plots}
\end{center}
\end{figure}

After performing the global fitting of all parameters, we find that the best-fit values of
the vanilla parameters are in agreement with the {\it Planck} values~\cite{Ade:2015xua}.
Then we present the posterior of $\xi$ and $k_c$ in Fig.~\ref{figure:1D_plots}.
Since the current data cannot provide any significant constraint on $c_H$, we do not report any
bound on it. We specifically consider the AMI with feature for the $N_\ast=50,~55,~60$,
and $N_\ast$ free cases. The improvement in the maximum likelihood obtained
in our runs with respect to $\Lambda$CDM is given by the quantity $\Delta\chi^2_{\mathrm{eff}}$,
which is found to be $-8.9,-11.6,-6.8,-6.2$, respectively.

\begin{figure}
\begin{center}
\includegraphics[scale=0.68]{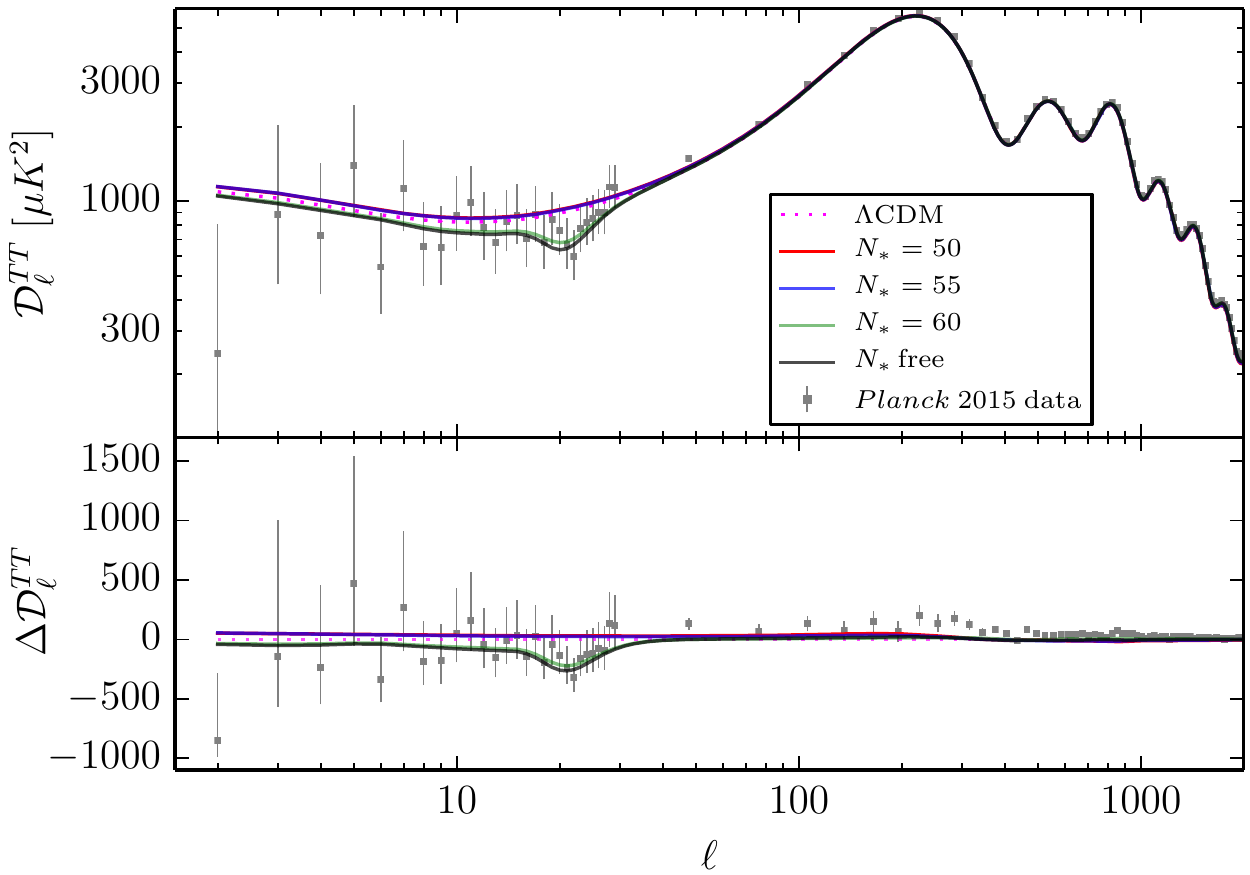}
\caption{The CMB temperature-temperature power spectra for the best-fit $\Lambda$CDM and AMI with feature models.
Additionally, we superimpose the {\it Planck} 2015 data. The vertical axis of the top panel is defined as
$\mathcal{D}_{\ell}^{TT} \equiv \ell(\ell+1) \mathcal{C}_{\ell}^{TT} /2\pi$.
For the bottom panel, we plot the power spectrum residual ($\Delta\mathcal{D}_{\ell}^{TT}$) with respect to the $\Lambda$CDM one.}
\label{bestCMB_TT}
\end{center}
\end{figure}

We find that $k_c\approx 0.0013\,{\rm Mpc}^{-1}$ is the best-fit value
of the scale at which the feature in the primordial power spectrum is located.
This roughly corresponds to the angular scale with multipole $\ell \approx 20$.
Indeed, we see from Fig.~\ref{bestCMB_TT} that for the $N_\ast=60$ and $N_\ast$ free cases,
the suppression feature becomes manifest near $\ell \approx 20$,
and accordingly, it leads to a slightly better fit to the {\it Planck} 2015 low-$\ell$ data~\cite{Aghanim:2015xee}.
Hence, our model can naturally explain the suppression anomaly near $\ell \approx 20$,
and the interpretation is that this observed anomaly is due to a short phase of slow-roll
violation during inflation~\cite{Cai:2014vua}.

We notice from Fig.~\ref{bestCMB_TT} that in the high-$\ell$ regime
our best-fit curves from {\it Planck}13 data have a slight
discrepancy with the {\it Planck} 2015 results. This is due to the fact that there
is roughly a $2\%$ shift in $A_se^{-2\tau}$ between the 2013~\cite{Ade:2013zuv}
and 2015~\cite{Ade:2015xua} data induced by the absolute calibration parameter correction
of the 143 GHz channel in the 2015 pipeline~\cite{Adam:2015vua}.
Moreover, our analysis reveals that the posterior of $\log_{10}k_c$ for the
$N_{\ast}=50$ and $55$ cases can have a second peak away from the anomaly scale,
hence they do not reproduce the $\ell\approx 20$ suppression in Fig.~\ref{bestCMB_TT}.
The anticorrelation between $N_{\ast}$ and $k_{\ast}$ (the larger $N_{\ast}$
is, the farther away from the end of inflation Hubble crossing occurs,
and hence the smaller $k_{\ast}$)
demonstrates that in the case of small $N_{\ast}$ (or large
$k_{\ast}$) one can fit the small scale structures in the primordial power
spectrum better and vice versa.

The current CMB data can only report an upper bound for the $\xi$ parameter,
$\log_{10}\xi \lesssim -1.2$ at $95\%$ C.L., and it is insensitive to the $c_H$ parameter compared to
its theoretical prior. As for the degeneracy between feature and $\Lambda$CDM parameters, no significant
correlation is obtained. The most interesting correlation is found between $k_c$ and the tensor-to-scalar
ratio ($r_{0.05}$). In the localized feature regime ($k_c\approx 0.0013\,{\rm Mpc}^{-1}$), the tensor-to-scalar
ratio is dragged to a greater value $r_{0.05}\lesssim 0.2$ at $95\%$ C.L.
This is because the potential modulation leads
to a power deficit in the scalar spectrum without affecting as much the tensor spectrum (see~\cite{Cai:2014vua}).

Several previous studies of CMB features reported $-\Delta \chi_{\mathrm{eff}}^2\sim \mathcal{O}(10)$.
However, these signals were not statistically conclusive
\cite{Benetti:2012wu, Planck:2013jfk, Ade:2015lrj, Benetti:2013cja, Hazra:2014goa, Achucarro:2013cva, Achucarro:2014msa, Hu:2014hra, Meerburg:2013cla, Meerburg:2013dla, Chen:2012ja, Chen:2014cwa}.
The improvement obtained here is also not statistically significant, but yet,
it shows that the localized feature is mildly favored by the data.
As pointed out in~\cite{Planck:2013jfk},
it is difficult to prove that these results are not coming from statistics overfitting of noisy data.
Moreover, one needs to be careful with the $\Delta \chi_{\mathrm{eff}}^2$ values found 
since the best-fit value provided by CosmoMC in its Metropolis-Hastings mode might not be fully trustworthy.

Observational signatures of the new model can also be analyzed within the CMB polarization data,
namely, the E-mode spectra. On the one hand, since oscillations in the temperature maps are strongly washed out on small
scales~\cite{Mortonson:2009qv}, the inclusion of the polarization maps will help us capture information of early
universe models at relevant scales. On the other hand, if the origin of the low-$\ell$ anomaly in the
{\it TT} spectrum comes from the inflationary phase, a similar signal should
also appear in the low-$\ell$ regime of the {\it TE} and {\it EE} spectra.

\begin{figure*}
\begin{center}
\includegraphics[scale=0.635]{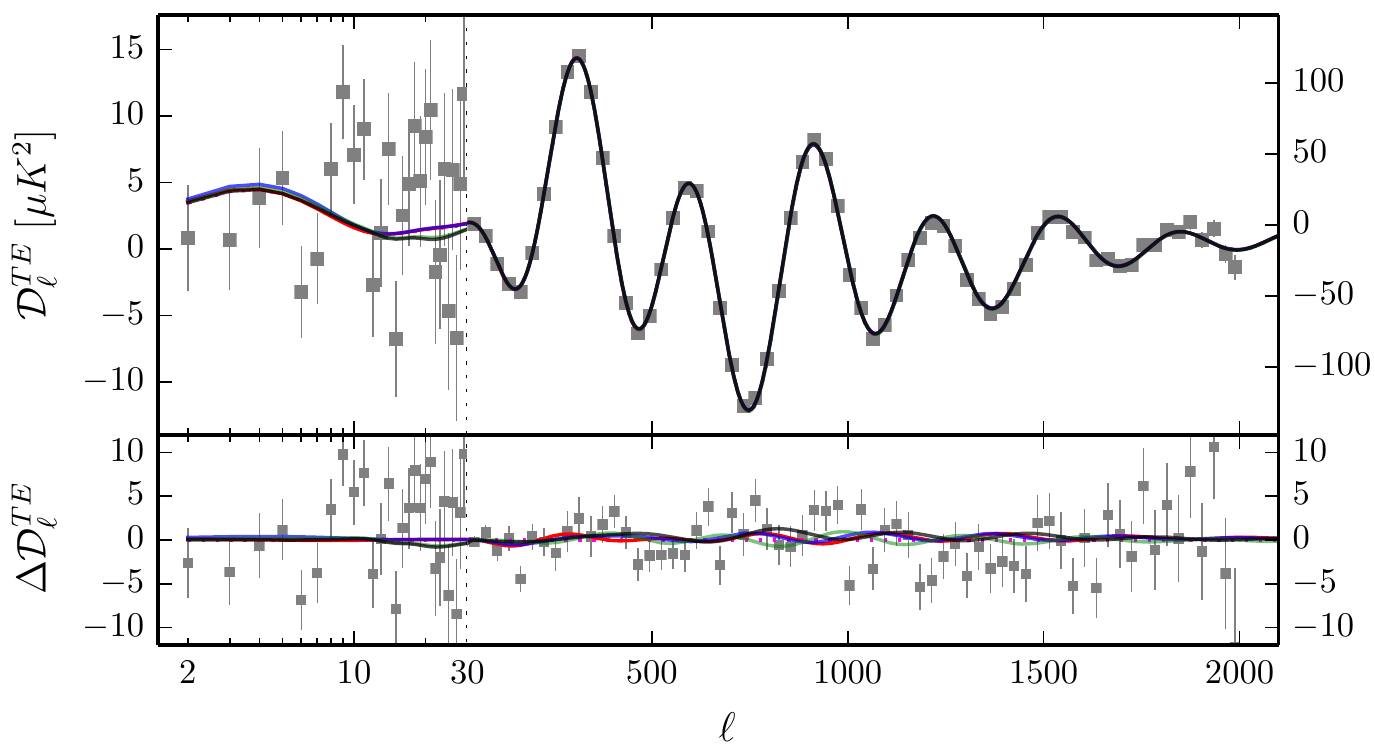}
\includegraphics[scale=0.635]{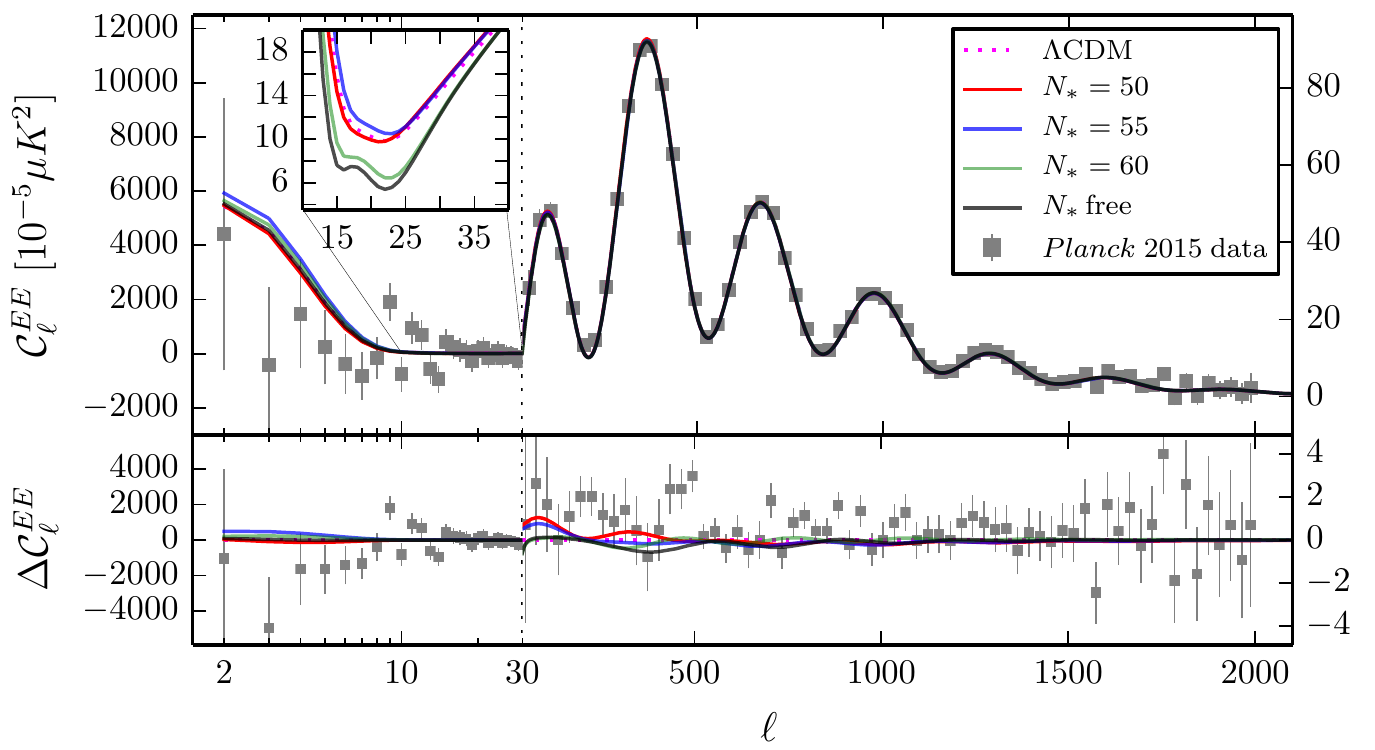}
\caption{Comparison between the theoretical CMB polarization power spectra ({\it TE} and {\it EE}) and the
{\it Planck} 2015 data. In the left panel, we show $\mathcal{D}_{\ell}^{TE}$ and the residuals
$\Delta\mathcal{D}_{\ell}^{TE}$ with respect to the $\Lambda$CDM best-fit. In the right panel, we show $\mathcal{C}_{\ell}^{EE}$ and
$\Delta\mathcal{C}_{\ell}^{EE}$. Note that the scales differ for $\ell<30$ and $\ell>30$ as shown by
the different vertical axis scales (the same way as Fig.~48 of~\cite{Aghanim:2015xee}).}
\label{bestCMB_TEEE}
\end{center}
\end{figure*}

The {\it Planck} collaboration happened to release the polarization data very recently~\cite{Aghanim:2015xee}.
In Fig.~\ref{bestCMB_TEEE}, we show the \textit{TE} and \textit{EE} power spectra for the
best-fit $\Lambda$CDM and AMI model with feature (obtained from the {\it Planck}13 data)
and we make a first comparison with the {\it Planck} 2015 data.
Looking at the residues of the \textit{TE} and \textit{EE} power spectra,
one can see that our model is distinct from $\Lambda$CDM both at
low-$\ell$ and high-$\ell$. At low-$\ell$, our model predicts power
deficits around $\ell \approx 15-30$ in the {\it TE} and {\it EE} spectra
(similar to the {\it TT} spectrum), of which the amplitudes
($\Delta \mathcal{D}_{\ell}^{TE}$ and $\Delta \mathcal{C}_{\ell}^{EE}$) are of order
$\mathcal{O}(1)[\mu{\rm K}^2]$ and $\mathcal{O}(5)[10^{-5}\mu{\rm K}^2]$, respectively;
while at high-$\ell$ the relative powers oscillate around $\Lambda$CDM in
both the {\it TE} and {\it EE} spectra. The oscillations at $\ell>30$ are
due to small differences in the best-fit $\Lambda$CDM parameters via the
mild correlations with the feature parameters, but the $\ell\approx 15-30$
suppressions in the {\it TE} and {\it EE} spectra are explicit predictions of the model
compared to $\Lambda$CDM. Although these suppression signals in the E-mode spectra
are difficult to be tested by the present observations since the differences
are currently much smaller than the overall scattering in the data, they
may provide a promising window for future CMB surveys which are expected to
greatly improve their accuracy of polarization measurements.

\section{Conclusions}

Evidence of fundamental theory (e.g.~string theory) occurs at extremely high energy scales,
and hence, is difficult to be found directly by experiments. To search for evidence, it is important to investigate the
associated cosmological implications, in particular, applications to the very early universe and their observational
consequences. In this sense, various theoretical models have been proposed to explain certain
CMB anomalies such as step potentials~\cite{Adams:2001vc, Benetti:2012wu, Adshead:2011jq, Miranda:2013wxa, Benetti:2013cja, Hazra:2014goa},
transient sound speed reduction~\cite{Achucarro:2010da, Achucarro:2013cva, Achucarro:2014msa, Hu:2014hra},
massive fields~\cite{Chen:2011zf, Chen:2012ja, Chen:2014cwa},
varying Planck mass models~\cite{Ashoorioon:2014yua},
pre-inflationary fast roll models~\cite{Gruppuso:2015xqa},
linear oscillations~\cite{Jackson:2013mka, Meerburg:2013cla, Meerburg:2013dla, Ashoorioonetal},
logarithmic oscillations~\cite{Martin:2000xs, Danielsson:2002kx, Bozza:2003pr},
and cutoff models~\cite{Contaldi:2003zv, Sinha:2005mn, Iqbal:2015tta}.

In the present paper, we studied the primordial power spectra of one class of inflationary model inspired by string
theory and examined their patterns in the CMB. We interestingly found a suppression feature in
the power spectrum of curvature perturbations in the vicinity of a critical scale, which could explain the power
deficit anomaly of temperature anisotropies as indicated by the recent {\it Planck} data.

We further studied the angular power spectra of CMB polarization modes and performed a preliminary
comparison with the {\it Planck} 2015 data. We showed that some patterns in these spectra are manifestly different from
$\Lambda$CDM and hence the model is observationally distinguishable if experimental accuracy is largely
improved in the future. While these polarization signals could be important,
it is also necessary to explore observational evidences from other avenues. For instance, we expect that the
modulation in the inflaton's potential can give rise to specific non-Gaussianity signals
(e.g., see~\cite{Achucarro:2012fd, Palma:2014hra, Novaes:2015uza, Mooij:2015cxa}) as well as signals
in the matter power spectrum that may be sensitive to the large scale structure experiments. The study of these
topics will be important along with this present work.

\begin{acknowledgments}
We thank Ana Achucarro, Frederico Arroja, Fang Chen, Chunshan Lin,
Daan Meerburg, Shinji Mukohyama, Misao Sasaki, Yi Wang, and Scott Watson for valuable discussions.
Y.F.C. is supported in part by the Chinese National Youth Thousand Talents Program and by the USTC start-up funding under Grant No.~KY2030000049.
E.F. thanks CNPq (Science without Borders) for financial support.
B.H. is supported by the Dutch Foundation for Fundamental Research on Matter (FOM).
J.Q. acknowledges the Fonds de recherche du Qu\'{e}bec - Nature et technologies (FRQNT) and
the Walter C. Sumner Foundation for financial support.
Computations were made in part on the supercomputer Guillimin from McGill University, managed by Calcul Qu\'{e}bec and Compute Canada.
The operation of this supercomputer is funded by the Canada Foundation for Innovation (CFI), NanoQu\'{e}bec, RMGA, and FRQNT.
\end{acknowledgments}

\end{document}